\documentclass[12pt]{article}
\usepackage{latexsym,epsf}
\newcommand{\beq}[3]{\begin{equation}  \label{#1#2#3}}
\newcommand{\eeq}{ \end{equation}}
\newcommand{\ba}{\begin{array}}
\newcommand{\ea}{\end{array}}

\newcommand{\remark}[1]{}
\let\LARGE=\Large
\let\Large=\large

\newcommand{\be}[3]{\begin{equation}  \label{#1#2#3}}     
\newcommand{\ee}{ \end{equation}}
\newcommand{\bea}{\begin{eqnarray}}
\newcommand{\eea}{\end{eqnarray}}

\newcommand{\ft}[2]{{\textstyle\frac{#1}{#2}}}

\def\beq{\begin{equation}}
\def\eeq{\end{equation}}
\def\beqa{\begin{eqnarray}}
\def\eeqa{\end{eqnarray}}

\begin{document}

\begin{titlepage}
\rightline{HU-EP-02/04}
\rightline{hep-th/0201270}

\vspace{15truemm}

\centerline{\bf \LARGE
Curved BPS domain walls 
and RG flow
}
%
\vspace{3mm}
\centerline{\bf \LARGE
in five dimensions}

\bigskip

\vspace{2truecm}

\centerline{\bf Gabriel Lopes Cardoso, Gianguido Dall'Agata
{\rm and} Dieter L\"ust}

\vspace{1truecm}

\centerline{\em  Institut f\"ur Physik, Humboldt University}
\centerline{\em Invalidenstra\ss{}e 110, 10115 Berlin, Germany}
\vspace{1truecm}
\centerline{\tt email:gcardoso,dallagat,luest@physik.hu-berlin.de}

\vspace{2truecm}


\begin{abstract}

We determine, in the context of five-dimensional ${\cal N}=2$ gauged
supergravity with vector and hypermultiplets, the conditions under which
curved (non Ricci flat) supersymmetric
domain wall solutions may exist.  These curved BPS domain wall solutions
may, in general, be supported by non-constant vector and hyper scalar
fields.  We establish our results by 
a careful analysis of the BPS equations 
as well as of the associated integrability conditions and 
the equations of motion.  We construct an example of 
a curved BPS solution in a gauged supergravity
model with one hypermultiplet.
We also discuss the dual description
of curved BPS domain walls in terms of RG flows.

\end{abstract}

\end{titlepage}


\newpage

\section{Introduction}

According to the 
domain wall/QFT correspondence \cite{BSK}, the renormalization group flow
in quantum field theories may be described in terms of a domain wall solution
in dual gauged supergravity theories.  An example of this is the RG flow
discussed by Freedman, Gubser, Pilch and Warner \cite{FGPW}.  This is a flow
in ${\cal N}=4$ super-Yang-Mills theory broken to an ${\cal N}=1$ theory
by the addition of a mass term for one of the three adjoint chiral superfields
\cite{KLM}.
Its dual description is in terms of a supersymmetric domain wall solution 
in five-dimensional ${\cal N}=8$ gauged supergravity which interpolates
between two $AdS$ vacua.  The ${\cal N}=2$ embedding of this UV-IR solution 
has been given in \cite{CDKV}.  The supersymmetric domain wall associated to
the flow of \cite{FGPW} is an example of a flat domain wall.  

In the context
of five-dimensional 
${\cal N}=2$ gauged supergravity, the study of BPS (i.e. supersymmetric)
domain walls has been mainly restricted to the case of Ricci flat
solutions \cite{CDKV}-\cite{BD}.  
It is, however, natural to ask whether 
there also exist curved (i.e. non Ricci flat) 
supersymmetric domain wall solutions.
Such domain wall solutions would presumably provide a dual gravitational
description of RG flows in supersymmetric field theories in a curved spacetime.
Another motivation for the study of curved domain wall solutions is given by
localized gravity on anti-de Sitter domain walls embedded in $AdS_5$
\cite{KR1}-\cite{BR}.

It is known \cite{CW,KK} that field theories and also string
theories in spacetimes with constant $AdS$
curvature exhibit improved infrared behaviour.  In a dual domain wall 
description, this improved infrared behaviour should come about by
turning on $AdS$ curvature on the domain wall.  In particular, since
in principle one may study the ${\cal N}=1$ super-Yang-Mills theory of 
\cite{FGPW} in a spacetime with constant $AdS$ curvature, one expects that 
it should be
possible to construct a curved (non Ricci flat) BPS domain wall solution
in ${\cal N}=2$ gauged supergravity which would provide a dual description 
of the RG flow.  Moreover, since curved domain wall solutions have
been constructed \cite{BCL} 
in four-dimensional ${\cal N}=2$ gauged supergravity, it 
is plausible that curved domain wall solutions also exist 
in five-dimensional 
${\cal N}=2$ gauged supergravity.  Thus, whereas UV divergences in 
four-dimensional
holographic
field theories are regulated by performing computations away from the boundary
of $AdS_5$ \cite{witten}, we expect that
IR divergences may be regulated by turning on 
curvature on the dual domain wall.

A first step towards the construction of curved 
supersymmetric domain wall
solutions in five dimensions was taken in \cite{CDL}.
There we considered five-dimensional ${\cal N}=2$ gauged
supergravity with vector and hypermultiplets, and we
analyzed the possibility of constructing curved BPS
domain wall solutions which are supported by non-constant scalar
fields.  The construction presented in \cite{CDL} is perfectly general
and may be applied to curved domain walls supported by scalar fields
belonging to only one type (vector or hyper)
or to both types of supermultiplets.
We showed that the resulting BPS equations for
the warp factor and for the vector scalars are modified by the presence
of a four-dimensional cosmological constant on the domain wall, extending
earlier results by DeWolfe, Freedman, Gubser and Karch 
\cite{DeWolfe:2000cp} in the context of non-supersymmetric gravitational
theories in five dimensions.
We also showed that the cosmological constant on the BPS domain wall must be
anti-de
Sitter like and that it constitutes an independent quantity, not related
to any of the objects appearing in the context of very special geometry.

Related work on curved  
domain walls appeared in \cite{CSabra}.
There it was argued that the integrability of the gravitini
equation rules out 
the existence of supersymmetric domain walls with a non-vanishing
cosmological constant on the wall.

Here we return to this issue and we show that curved BPS domain wall
solutions may very well exist in five-dimensional ${\cal N}=2$ gauged
supergravity with vector and hypermultiplets.
We do this by first analyzing
the BPS flow equation for the hyper scalars, which we didn't give
in \cite{CDL}.
We then use this information to 
check whether the integrability conditions derived from 
the gravitini equation associated to 
curved BPS domain wall solutions
are satisfied, and we establish that this is indeed the case.

We then give the energy functional for curved domain wall solutions.
Interestingly we find that it isn't just given in terms of squares of 
BPS equations and of boundary terms, but that there are also contributions
that are linear in the BPS equations as well as an additional term
proportional to the warp factor.  The latter, whose presence was already
noted in 
\cite{DeWolfe:2000cp}, is crucial for 
ensuring that curved solutions to the BPS flow equations also
solve the Einstein equations of motion.

We use the energy functional to compute the equations of motion for the
vector and the hyper scalars.
The resulting equations 
are complicated.  It is not guaranteed that a solution to the BPS 
flow equations 
will automatically solve the equations of motion for the scalar
fields, as was already pointed out in \cite{CSabra}.  This is tied to the
presence, in the energy functional, of terms that are linear in the BPS
equations.  Thus, in order to construct curved BPS domain wall solutions
one should proceed as follows.
First
one solves the flow equations for the warp factor and for 
the scalar fields.  
Then 
one plugs the resulting expressions into the equations of motion for the
scalar fields and checks whether they are satisfied.  
Note that already in the case of flat domain walls the equations of
motion yield an additional condition \cite{CDKV} which must be fullfilled by 
a solution to the BPS flow
equations.

We then discuss the RG flow interpretation of curved BPS
domain wall solutions.  

And finally, we give an example of a curved BPS domain wall solution.
We explicitly construct such a 
solution in a gauged supergravity model with one hypermultiplet
only, 
and we check that the equations of motion are satisfied.

\section{Five-dimensional curved BPS domain wall solutions}

The five-dimensional ${\cal N}=2$ gauged supergravity theories that
we consider are in the class constructed in
\cite{AnnaGianguido}, describing the general coupling of
$n_V$ vector multiplets and of $n_H$
hypermultiplets to supergravity.
The scalar fields $\phi^x$ ($x=1, \dots, n_V$) of the vector multiplets
parametrize a very special manifold.
The hypermultiplet scalars $q^X$, on the other hand, parametrize a
quaternionic K\"{a}hler geometry determined by $4n_H$-beins
$f^{iA}_X(q^X)$, with the $SU(2)$ index $i=1,2$ and
the $Sp(2n_H)$ index $A=1,\ldots,2n_H$, raised and lowered by the
symplectic
metrics $\varepsilon_{ij}$ and $C_{AB}$. 
We refer to \cite{AnnaGianguido} for more details.

The scalar potential in such theories
is given by \cite{CDKV}
\begin{equation}
 {\cal V}= - 6 \, W^2+\frac{9}{2} \, g^{\Lambda\Sigma}\partial_\Lambda
 W\partial_\Sigma W\, + \frac{9}{2}\,  W^{2}
 (\partial _x Q^s)(\partial ^x Q^s) \;. \label{pot}
\end{equation}
Here $g_{\Lambda \Sigma}$ denotes the metric of the complete scalar
manifold, which is positive definite,
involving the scalars of both the vector and the hypermultiplets.
$W(\phi,q)$ and 
$Q^s(\phi,q)$ denote the norm and the $SU(2)$ phases, respectively,
which  appear in the decomposition
of the triplet of Killing prepotentials $P^s$, i.e.
$P^s = \sqrt{\ft{3}{2}} W Q^s$ with $ Q^s Q^s = 1$.
Note that in (\ref{pot}), 
the derivatives acting on  the $SU(2)$ phases $Q^s$ are only computed
with respect to the scalars of the vector multiplets.  

It will be
convenient to rewrite (\ref{pot}) as follows,
\begin{equation}
 {\cal V}= - 6 \, W^2+\frac{9}{2} \, \Gamma^{-2}\, g^{x y}\partial_x
 W\partial_y W\, 
+\frac{9}{2} \, g^{XY}\partial_X
 W\partial_Y W\, 
\;, \label{pot2}
\end{equation}
where
\bea
\Gamma^{-2} (\phi,q) = 
1 + W^2 \, \frac{g^{xy} (\partial _x Q^s)(\partial_y Q^s)}{
 g^{x y}\partial_x
 W\partial_y W } \;.
\label{Gamma}
\eea
We are interested in the construction of curved BPS domain wall solutions.
These are solutions which are uncharged, which are supported by non-constant
scalar fields and which have residual supersymmetry.  We take the associated
spacetime metric to be given by
\be100
ds^2 = {\rm e}^{2 U(r)} {\hat g}_{mn}
dx^m dx^n + 
dr^2 \;,
\label{line}
\ee
where $x^{\mu} = (x^m, r)$ (with $x^m = (t,x,y,z)$)
 and ${\hat g}_{mn} = {\hat g}_{mn} (x^m)$.  
We denote the 
corresponding tangent space indices by $a = (0,1,2,3,5)$.  
The metric ${\hat g}_{mn}$ is taken to be a four-dimensional constant curvature
metric, i.e. $
{\hat R}_{mn} = - 12 l^{-2} \, {\hat g}_{mn} 
$, 
with the four-dimensional cosmological constant proportional to $l^{-2}$.
For the solution to be supersymmetric, we must take $l$ to be 
imaginary \cite{CDL},
which corresponds to a four-dimensional
anti-de Sitter spacetime.
Since these solutions are uncharged, we set the
gauge fields to zero.\footnote{It can be checked that the equations of
motion for the gauge fields are satisfied by these solutions.}  
We allow for a non-trivial dependence of the scalar
fields on the coordinate $r$, and we write $\phi' = d \phi /d
r$, $q'= d q / d r$ as well as $U' = d U / d r$.

A solution with residual supersymmetry 
is obtained
by requiring that the supersymmetry transformation laws,
when evaluated on a given solution,
 vanish
for some combination of the supersymmetry transformation parameters
$\epsilon_i$.  For solutions with spacetime line element (\ref{line}),
the appropriate 
projector condition on the supersymmetry transformation parameters
$\epsilon_i$ leading to residual supersymmetry is given by \cite{CDL}
\bea
i \gamma_5 \epsilon_i = A(r) \; {Q_i}^j \epsilon_j + B(r) \; {M_i}^j
\epsilon_j \;.
\label{procmod}
\eea
Here $Q = i Q^s \sigma^s$ and $M = i M^s \sigma^s$
denote $SU(2)$-valued matrices 
satisfying $Q_i\,^j Q_j\,^k =- \delta_i\,^k , M_i\,^j M_j\,^k =- \delta_i\,^k$
(i.e. $Q^s Q^s = 1 , M^s M^s =1$).
Without loss of generality, we take $Q$ and $M$ to be orthogonal in $SU(2)$
space, so that $Q^s M^s = 0$.
The consistency of (\ref{procmod}) then yields that
\bea
A^2(r) + B^2(r) = 1 \;.\label{AB}
\eea
A curved BPS domain wall solution is supported by a non-trivial warp factor
$U$ as well as by non-constant scalar fields.  For the solution to be 
supersymmetric, these various fields have to satisfy a set of so-called
BPS flow equations.
The flow equation for the warp factor $U$ can be derived as follows \cite{CDL}.
Demanding the vanishing of the gravitini variation 
$\delta \psi_{mi} $
and inserting
(\ref{procmod}) into it yields 
\bea
\hat{\cal D}_m \epsilon_i = \ft{i}{2} \left( U^\prime \, A\, + g W
\right) \gamma_m {Q_i}^j \epsilon_j + \ft{i}{2} U^\prime B
\gamma_m  {M_i}^j
\epsilon_j \;,
\label{gravm}
\eea
whose integrability gives 
\bea
2 U^\prime \left( U^\prime  + A g W  \right) = (U^\prime)^2 - 4
l^{-2} {\rm e}^{-2U} - g^2 W^2\;.
\label{relab}
\eea
On the other hand, the integrability of $\delta \psi_{\mu i} $ also
gives rise to 
\bea
(U^\prime)^2 - 4 l^{-2}\, {\rm e}^{-2U} -  g^2 W^2 = 0\,, \label{eins2}
\eea
which yields
\bea
U' = \pm \gamma(r) \, g W \;,
\label{modwarp}
\eea
where
\bea
\gamma(r) \equiv \sqrt{1 - \frac{4 {\rm e}^{-2U}}{ |l|^2 g^2W^2}}\;.
\label{gamma}
\eea
Combining (\ref{relab}) 
and (\ref{eins2}), however,
also gives
\bea
U^\prime  = -  A g W  \;.
\label{flowu}
\eea
Then, by comparing (\ref{modwarp}) with (\ref{flowu}) 
we obtain $A$ as a function of
$W$ and of $U$, namely
\bea
A = \mp \gamma (r) \;.
\label{ag}
\eea
The flow equations for the scalar fields belonging to the vector
multiplets is derived as follows \cite{CDL}.
Demanding the vanishing of the supersymmetry variation of the gaugini
subject to (\ref{procmod}) yields 
\begin{equation}
A(r) \phi^{x \,\prime} =  3 g\, g^{xy} \partial_{y} W
\label{dWvector}
\end{equation}
as well as 
\begin{equation}
B(r) M^s \phi^{x \prime} = 3 \,g W \; g^{xy} \partial_{y} Q^s\,.
\label{dQr0}
\end{equation}
Combining (\ref{dWvector}) and (\ref{dQr0}) we obtain
\bea
A^{-2}  = 
1 + W^2 \, \frac{g^{xy} (\partial _x Q^s)(\partial_y Q^s)}{
 g^{x y}\partial_x
 W\partial_y W } \;.
\eea
Inspection of (\ref{Gamma}) then gives another expression for $A$ in terms
of the scalar fields, namely
\bea
A = \mp \Gamma \;.
\label{aG}
\eea
The flow equation 
for the
vector scalars thus reads
\begin{equation}
\phi^{x \,\prime} =  \mp 3g \, \Gamma^{-1} \, g^{xy} \partial_{y} W\,.
\label{flowv}
\end{equation}
Combining (\ref{dQr0}) and (\ref{flowv}) yields
\bea
\partial_x Q^s = \mp B \Gamma^{-1} W^{-1} \partial_x W \, M^s\;.
\label{parq}
\eea
By squaring (\ref{parq}) we precisely obtain (\ref{Gamma}).

Comparing (\ref{ag}) with (\ref{aG}) yields
\bea
\Gamma (\phi, q) = \gamma (r) \;.
\label{cons1}
\eea
Note that (\ref{cons1}) only applies if a given gauged supergravity model
has vector multiplets, since both the definition (\ref{Gamma}) and 
the derivation of (\ref{aG}) depend on that.  
For models with vector multiplets,
(\ref{cons1}) constitutes a consistency check on the solution,
since $\Gamma (\phi, q)$ only depends on the scalar fields, whereas
$\gamma (r)$ is computed
from both the warp factor and the scalar fields.  
We observe that (\ref{cons1}) is solved by
$\partial_x Q^s = \mp B \gamma^{-1} W^{-1} \partial_x W \, M^s$, 
which is nothing but (\ref{parq}).
The case
of flat domain walls corresponds to $\gamma =1$, $B=0$ and
hence $\partial_x Q^s =0$.

The flow equations (\ref{modwarp}) and (\ref{flowv}) 
were also derived in \cite{DeWolfe:2000cp} in the context of
non-supersymmetric five-dimensional gravity theories with a single
scalar field.

Let us now turn to the flow equation for the scalar fields belonging
to the hypermultiplets, 
which we didn't analyze in \cite{CDL}.  The vanishing of the hyperini
equations, when subjected to (\ref{procmod}), yields the following
flow equation for the hyper scalars,
\bea
\ft{1}{3} g_{XY} \, q^{Y \prime} = g A \partial_X W + g W B M^s D_X Q^s \;,
\label{flowh}
\eea
which may also be rewritten as \cite{CSabra}
\bea
g_{XY} \, q^{Y \prime} = 3 g {\Sigma_X}^Y \partial_Y W \;\;\;,\;\;
{\Sigma_X}^Y = A {\delta_X}^Y + 2 B \varepsilon^{rst} M^r Q^s {R^t_X}^Y \;,
\label{flowhs}
\eea
where ${R^t_X}^Y$ denotes the $SU(2)$ curvatures.
Moreover, it can be checked that
\bea
{\Sigma_X}^Y g^{XZ} {\Sigma_Z}^V = g^{YV} \;.
\label{ss}
\eea
Then, using (\ref{flowhs}) and (\ref{ss}), it follows that
\bea
 g_{XY} \, q^{X\prime} q^{Y\prime} = 9 \, g^2 
g^{XY} \, \partial_X W \partial_Y W \;.
\label{qq}
\eea
On the other hand, contracting (\ref{flowhs}) with $\partial^X W$
yields
\bea
q^{X\prime} \partial_X W = 3 g A g^{XY} \partial_X W  \partial_Y W \;.
\label{qw}
\eea
Comparing (\ref{qq}) and (\ref{qw}) then gives
\bea
g_{XY} \, q^{X \prime} \, q^{Y \prime} = 3 g A^{-1} 
 q^{X \prime} \, \partial_X W =
\mp 3 g \gamma^{-1}
 q^{X \prime} \, \partial_X W \;,
\label{flowhc}
\eea
where we used (\ref{ag}).  Thus, we observe that the contracted version 
(\ref{flowhc}) of the flow equation for the hyper scalars has a similar
structure as the contracted version of the flow equation for the 
vector scalars (\ref{flowv}) given by $g_{xy} 
\phi^{x \,\prime} \phi^{y \,\prime}
=  \mp 3g \, \gamma^{-1} \, \phi^{x \,\prime}  \partial_{x} W$.

Another useful relation can be obtained as follows.  Contracting 
(\ref{flowh}) with $q^{X \prime}$ and using (\ref{flowhc}) yields
\bea
q^{X \prime} 
B M^s
\, D_X  Q^s = \mp 
\frac{(1 - \gamma^2)}{\gamma W } \, q^{X \prime} \, \partial_{X} W \;.
\label{cons2}
\eea
Now, by using 
$D_r  Q^s = \phi^{x \prime} \, \partial_{x} Q^s + q^{X \prime} 
D_X Q^s$ and 
$W' = \phi^{x \prime} \, \partial_{x} W + q^{X \prime} \partial_X W $ 
as well as 
(\ref{parq}), (\ref{cons1}) and (\ref{cons2}), we obtain
\bea
B M^s
\, D_r  Q^s = \mp
 \frac{(1 - \gamma^2)}{\gamma W } \, W^{\prime} \;.
\label{cons7}
\eea
This concludes our discussion of the derivation 
of the BPS flow equations for the
various fields supporting a curved BPS domain wall.

We proceed to check the various integrability conditions associated to
$\delta \psi_{\mu i} =0$.  One such condition is given by (\ref{eins2}).
Another is 
given by \cite{CDL}
\bea
3 U^{''} + 12 l^{-2}\, {\rm e}^{-2U} = - g_{xy} \, 
\phi^{x\prime}\phi^{y\prime}
-  \ft12 g_{XY} \, q^{X\prime} q^{Y\prime}  - \ft92 \, g^2 
g^{XY} \, \partial_X W \partial_Y W \;.
\label{eins1}
\eea
By 
inserting (\ref{modwarp}), (\ref{flowv}), (\ref{qq}) and (\ref{flowhc})
into (\ref{eins1}) we find that (\ref{eins1}) is identically satisfied.

As pointed out in \cite{CSabra}, there is one more integrability condition
that needs to be checked, namely $[{D}_r , {\hat {\cal D}}_m] \epsilon_i
=0$.  To evaluate this, we use
the gravitini variation equation
$\delta \psi_{ri} =0$ which results in
\bea
D_r \epsilon_i = \ft{i}{2} g W Q_i\,^j \gamma_5 \epsilon_j \,,
\label{gr}
\eea
where $D_r \epsilon_i = \partial_r 
\epsilon_i - q^{X'} \omega_{Xi}\,^j \epsilon_j $.  
Inserting the projector
condition (\ref{procmod})
into the rhs of (\ref{gr}) then yields
\bea
D_r \epsilon_i = - \ft{1}{2} g W A \epsilon_i + \ft{1}{2} g W B (QM)_i\,^j 
\epsilon_j \;.
\label{gr2}
\eea
Using 
(\ref{gravm}) as well as (\ref{gr2}) we now evaluate 
$[{D}_r , {\hat {\cal D}}_m] \epsilon_i
=0$ and 
obtain\footnote{Our expressions
differ from the ones given in \cite{CSabra}.}
\bea
\pm \gamma^{\prime} + B M^s \, D_r Q^s = B^2 g W 
\label{intl1}
\eea
as well as
\bea
\pm B^{\prime} + \gamma Q^s \, D_r M^s = - B \gamma g W \;.
\label{intl2}
\eea
It is easy to check that (\ref{intl1}) implies (\ref{intl2})
and vice-versa.  Now, using (\ref{gamma}) 
we compute $\gamma^{\prime} = B^2 [\pm g W + 
\ft{W^{\prime}}{\gamma W}]$.  By inserting this as well as (\ref{cons7}) into
(\ref{intl1}), we find that (\ref{intl1}) is identically satisfied.
Thus, similarly to the case of flat domain walls, 
there are no additional requirements on a curved domain wall solution
steming from the integrability conditions associated with 
 $\delta \psi_{\mu i} =0$.

Next we derive the energy functional for curved domain wall solutions
(\ref{line}).  We will subsequently use it to compute the equations of motion
for the various fields supporting curved BPS domain walls.

The energy functional $E$ for non-trivial solutions of the form
(\ref{line}) may be written as follows.  Denoting $V_4 = \int \sqrt{|\det
{\hat g}_{mn}|}$, we can rewrite the bulk action $S_{\rm bulk}$ as follows,
\bea
E/V_4 \propto S_{\rm bulk}/V_4 &=& \int dr \, {\rm e}^{4U} \Big[
\ft{1}{2}
(\phi^{x \prime} \pm 3  \Gamma^{-1} g \partial^x W)^2
+ \ft{1}{2} (q^{X \prime} - 3 g \Sigma^{XY} \partial_Y W)^2 
\nonumber\\
&& \;\;\;\;\;\;\;\;\;\;\;\;\;\; 
- 6 (U^\prime \mp \gamma g W)^2  \Big]  \nonumber\\
&+& 3g \int dr \, {\rm e}^{4U} \Big[ \mp \phi^{x \prime} \partial_x W (
\Gamma^{-1} - \gamma^{-1}) + q^{X \prime} ( {\Sigma_X}^Y \partial_Y W
\pm \gamma^{-1} \partial_X W) \Big]
\nonumber\\
&\pm&  \ft{12}{|l|^2} \int dr \, {\rm e}^{2U} 
\left[ \frac{U^\prime}{\gamma g W} -1 \right]
+ \int dr \, \frac{d}{dr} \Big[ 4 {\rm e}^{4U} U^\prime 
\mp 3 {\rm e}^{4U} \gamma g W
\Big]
 \nonumber\\
& \pm & \ft{12}{|l|^2} \int dr \, {\rm e}^{2U}  \;
\;.
\label{energy}
\eea
Thus we see that the energy functional for curved walls is not just 
given in terms of squares of BPS equations and of total derivatives
(as it is the case for flat domain wall solutions \cite{ST}), but that
there are also contributions 
that are linear in the BPS equations as well as an
additional term (the last term) proportional to the warp factor.
The latter, as noted in \cite{DeWolfe:2000cp}, cannot be rewritten
into BPS equations and total derivative terms.  Its presence is, however,
crucial for ensuring that curved solutions to the BPS flow equations also
solve the Einstein equations of motion given by 
\bea
4 U^{\prime \prime} + 4 (U^\prime)^2 &=& - g_{\Lambda \Sigma} \phi^{\prime 
\Lambda} \phi^{\prime 
\Sigma} - \ft23 g^2 {\cal V} \;, \nonumber\\
U^{\prime \prime} + 4 (U^\prime)^2 + 12 |l|^{-2} {\rm e}^{-2U} 
&=& - \ft23 g^2 {\cal V} \;.
\eea
Indeed, varying (\ref{energy}) with 
respect to $U$ yields that $\delta_U E = 0 $ when evaluated on a curved
BPS solution satisfying (\ref{modwarp}),
(\ref{flowv}), (\ref{cons1}),
(\ref{flowhs}) and (\ref{flowhc}).  

Now consider varying (\ref{energy}) with respect to a vector scalar field
$\phi^x$.  Demanding that $\delta_x E =0$ on a curved BPS solution, we obtain
\bea
\pm \Big( \phi^{y \prime} \partial_y W  \, \partial_x \Gamma - \gamma^{\prime}
\partial_x W \Big) + \gamma^2 q^{X \prime} \Big( {\Sigma_X}^Y 
\partial_x \partial_Y W \pm \gamma^{-1} \partial_x \partial_X W 
\Big) =0 \; ,
\label{mov}
\eea
where we used $\Gamma^\prime = \gamma^\prime$ as well as
$q^{X \prime} (\partial_x {\Sigma_X}^Y) \partial_Y W =0$
by virtue of (\ref{flowhs}).
On the other hand, varying (\ref{energy})
with respect to a hyper scalar $q^Z$ and demanding that 
$\delta_Z E =0$ on a curved BPS solution yields 
\bea
\label{moh}
&&\pm \gamma^{-2}  \phi^{x \prime} \partial_x W  \, \partial_Z \Gamma 
\mp 4 g \gamma W {\Sigma_Z}^Y \partial_Y W - 4 g W \partial_Z W \\
&& + q^{X \prime} (\partial_Z {\Sigma_X}^Y ) \partial_Y W 
- {\Sigma^\prime_Z}^Y \partial_Y W \nonumber\\
&& - q^{X \prime} \Big( {\Sigma_Z}^Y \partial_X \partial_Y W
- {\Sigma_X}^Y \partial_Z \partial_Y W \Big) 
- \phi^{x \prime} \Big( \pm \gamma^{-1} \partial_Z \partial_x W +
{\Sigma_Z}^Y \partial_Y \partial_x W \Big) = 0 \;. \nonumber
\eea
We note that if we contract (\ref{mov}) with $\phi^{x \prime}$
and (\ref{moh}) with $\gamma^2 q^{Z \prime}$,  and if we use that $
{\Sigma_X}^Y = {\Sigma_X}^Y (U, \phi, q)$,
then the sum of the
resulting equations vanishes identically.

It can be checked that (\ref{mov}) and (\ref{moh}) are indeed
nothing but the equations of motion for the scalar fields
\bea
{\rm e}^{-4U} ({\rm e}^{4U} g_{\Lambda \Sigma} \phi^{\prime \Sigma })^{\prime}
- \ft12 (\partial_{\Lambda} g_{\Gamma \Sigma}) \,  \phi^{\prime \Gamma}
\phi^{\prime \Sigma} = g^2
\partial_{\Lambda} {\cal V} \;
\eea
when evaluated on a curved BPS domain wall solution.

If a given gauged supergravity model contains only hypermultiplets,
then (\ref{mov}) is trivially satisfied, whereas (\ref{moh})
constitutes a consistency check for a given solution to the BPS
flow equations (\ref{modwarp}) and (\ref{flowhs}).  If, on the other hand,
a gauged supergravity model contains vector multiplets (with or without
hypermultiplets), then (\ref{mov}) and (\ref{moh}) may be viewed as
equations which determine the second derivatives $\partial_x \partial_y
Q^s$ and $\partial_x \partial_X Q^s$ through
$\partial_x \Gamma$ and $\partial_X \Gamma$.  These expressions
will then have to agree with the ones obtained by inserting a solution
to the BPS flow equations into
$\partial_x \Gamma$ and $\partial_X \Gamma$.  Otherwise a solution to
the BPS flow equations will not solve the equations of motion for the
scalar fields.  In the case of flat domain walls it follows from 
(\ref{mov}) and (\ref{moh}) that $\partial_x \Gamma = \partial_X \Gamma =0$,
which is in accordance with $\partial_x Q^s =0$ as derived from
(\ref{parq}).

\section{Curved BPS domain walls and RG flow}

Let us briefly summarise the properties of curved BPS domain wall 
solutions that we have constructed.  
The BPS flow equations for the warp factor and for the scalar fields
are given by
\bea
U^\prime &=& \pm \gamma g W \;,
 \nonumber\\
\phi^{x \,\prime} &=&  \mp 3g\, \gamma^{-1} \, g^{xy} \partial_{y} W\;,
\nonumber\\
q^{X \prime} &=& 3 g\, g^{XY} {\Sigma_Y}^Z \partial_Z W \;,
\label{floweqs}
\eea
where $
\gamma(r)=\sqrt{1 - 4 |l|^{-2} ({\rm e}^{U} g W)^{-2}}$ and where 
${\Sigma_X}^Y = A {\delta_X}^Y + 2 B \varepsilon^{rst} M^r Q^s {R^t_X}^Y.$
Here $A = \mp \gamma \;,\; A^2 + B^2 =1,$ and ${\Sigma_X}^Y$ satisfies
${\Sigma_X}^Y g^{XZ} {\Sigma_Z}^V = g^{YV}$.
In the case that a given gauged supergravity model contains vector multiplets,
a solution to the BPS flow equations (\ref{floweqs}) 
will also have to satisfy the following 
consistency condition due to supersymmetry, 
\bea
\Gamma (\phi, q) = \gamma (r) \;,
\eea
or equivalently,
\bea
\partial_x Q^s = \mp B \gamma^{-1} W^{-1} \partial_x W \, M^s \;.
\label{pq}
\eea
If a gauged supergravity model contains only hypermultiplets, then this
consistency condition is absent.

It is not guaranteed that a solution to the BPS flow equations (\ref{floweqs})
will also automatically solve the 
scalar equations of motion (\ref{mov}) and (\ref{moh}).  Thus, in order to
construct a curved BPS domain wall solution one should proceed as follows.
First one constructs a solution to the BPS flow equations
(\ref{floweqs}).  In the presence of hypermultiplets, this requires choosing
a triplet $M^s$ such that $M^s Q^s = 0 , M^s M^s =1$.
It may be that in order to explicitly construct a solution to (\ref{floweqs}),
one has to expand $\gamma$ and ${\Sigma_X}^Y$ 
in powers of $|l|^{-1}$ and solve the flow equations iteratively as a power
series in $|l|^{-1}$.
Then one checks whether the consistency condition
(\ref{pq}) is met by this solution (this
only applies to models which include vector multiplets).
If so, then one finally checks
the scalar equations of motion (\ref{mov}) and (\ref{moh}).  If they are
satisfied, one has managed to construct a curved BPS domain wall solution.

Observe that, on a solution to the BPS flow equations (\ref{floweqs}),
the potential (\ref{pot2}) may be written in a more symmetric way,
namely as
\bea
 {\cal V} &=& - 6 \, W^2+\frac{9}{2} \, \gamma^{-2}\, g^{xy} \partial_x
 W\partial_y W\, 
+\frac{9}{2} \, {\Sigma_Z}^X g^{ZV} {\Sigma_V}^Y 
\partial_X
 W\partial_Y W\, \nonumber\\
&=& - 6 \, W^2  + \frac{1}{2} \, g_{xy} \, \phi^{x \prime}
\phi^{y \prime} + \frac{1}{2} \, g_{XY} \, q^{X \prime} q^{Y \prime}
\;. \label{potsol}
\eea

Let us now turn to the RG flow interpretation of a curved BPS domain 
wall solution.
In the case of flat domain walls with line element $ds^2 = {\rm e}^{2U} 
\eta_{mn}
dx^m dx^n + dr^2$, 
the renormalisation group scale $\mu$ of the dual field theory 
is usually identified with $\mu = {\rm e}^U$ \cite{FGPW}, 
where $0 < \mu < \infty$.  The field theory UV region ($U \rightarrow \infty, 
\mu \rightarrow \infty $) is identified with $r \rightarrow \infty$.
This identification amounts to choosing the upper sign in the flow
equations (\ref{floweqs}).  
A flow towards the infrared in field theory 
($U \rightarrow - \infty \,,\, \mu \rightarrow 0 $)
then corresponds to a 
a flow towards smaller values of $r$.

In the case of curved domain 
walls, we again identify ${\rm e}^U$ with the 
renormalisation group scale $\mu$ of the dual field theory, $\mu={\rm e}^U$.
Now, however, 
it may happen that $U$ cannot run anylonger 
over the whole range $-\infty < U  < \infty$.  This is tied to the fact
that $\gamma$, which can have a value between 
$0 \leq \gamma \leq 1$, becomes vanishing whenever
${\rm e}^{U} = \frac{2}{|l|} (gW)^{-1}$.
Since both $W$ and $U$ are functions of $r$, 
this will happen at 
specific values of $r$.  
In order to discuss the various possibilities,
let us introduce $T = ({\rm e}^{U} g W)^{-1}$ and $\Lambda^2 = 4 |l|^{-2}$,
so that $\gamma (r) = \sqrt{1 - \Lambda^2 T^2 (r)}$.  Let us then
consider deforming a flat domain wall
solution (which has $\gamma =1$) by 
turning on 
$\Lambda^2$ (which corresponds to turning on  
curvature on the domain wall).  There are then two possibilities which we will
now discuss.

The first possibility consists in the following.
When turning on a small amount $\Lambda^2$, 
$\gamma (r)$ remains non-vanishing along the flow.  That is, 
for small values of $\Lambda^2$, $T^2(r)$ has
the property that $T^2(r) < \Lambda^{-2}$ along the curved solution.
This curved solution exists over the whole range of $r$.
Then, when increasing $\Lambda^2$, there will
be a critical value $\Lambda^2_c$ at which 
$T^2(r_c) = \Lambda_c^{-2}$, i.e.  $\gamma (r_c) =0$ at
a certain position $r=r_c$.  
At this critical value of $\Lambda^2$, there is yet no obstruction in 
constructing a curved solution over the whole range of $r$.
Now let us continue to increase the value of $\Lambda^2$ (corresponding
to a large amount of curvature on the wall).  
Then $T^2(r) > \Lambda^{-2}$ in a certain region of $r$, i.e.
$\gamma$ becomes
imaginary in that region.  This region will be delimited by
the zeros of $\gamma$.  In this region there is no real solution to the
curved BPS equations.  The curved domain wall solution can only
be constructed outside the region where $\gamma$ is becoming imaginary.
The associated warp factor will now only cover part of the range
$-\infty < U  < \infty$.   
As we continue to increase the value of $\Lambda^2$, the forbidden
region becomes larger and larger, and the region where the solution 
exists smaller and smaller.  An example of a curved domain wall solution
exhibiting these features will be given in section \ref{secexa}.

The second possibility is the following.  In contrast to the one discussed
above, 
now, as soon as one turns on a small amount of
curvature on the wall, $\gamma$ develops a zero somewhere, 
i.e. $\gamma (r_{IR})=0$ at $T^2(r_{IR}) = \Lambda^{-2}$.
Let us assume that $T(r)$ is a monotonic function.  
Then the value $r_{IR}$ delimits two regions.  In the range $r_{IR} \leq r <
\infty$, $\gamma$ has the property that  
$\gamma \geq 0$, whereas for values $r< r_{IR}$ $\gamma$ becomes imaginary.
A real solution to the curved BPS equations then only exists in
the range $r_{IR} \leq r <
\infty$.  The warp factor $U$ runs over the range
$U_{IR} \leq U < \infty$, with $U_{IR}$ determined by 
$U_{IR} = \log[\Lambda (gW(r_{IR}))^{-1}]$.
Thus, by identifying ${\rm e}^U$ with an RG energy
scale $\mu$, we see that a non-vanishing four-dimensional
cosmological constant $\Lambda^{2}$
acts as an infrared cutoff in the field theory, i.e.
$\mu_{IR} = {\rm e}^{U_{IR}} \leq \mu < \infty$.
A curved domain wall then provides a dual
gravitational description of an RG flow 
in a field theory on an $AdS_4$ space, with its
curvature acting as an infrared regulator \cite{CW}.  
Observe that, as $\Lambda \rightarrow \infty$,
$\mu_{IR} \rightarrow \infty$ (provided that $W$ remains finite).  
Cranking up the curvature on the
domain wall thus pushes the infrared regulator towards the UV region.

Let us now proceed and show that a c-function also exists for curved
BPS domain wall solutions.
Using the flow equations (\ref{floweqs}) as well as (\ref{flowhc})
we compute 
\bea
U'' - 4|l|^{-2} {\rm e}^{-2U} 
= \gamma^{-1} W^\prime = - 3 
g^{\Lambda \Sigma}
\partial_{\Lambda} W \partial_{\Sigma} W - 3 (\gamma^{-2} -1) \, g^{xy}
\partial_x W \partial_y W \leq 0 \;,
\label{c}
\eea
where $\Lambda, \Sigma$ run both over the vector and the hyper scalars.
Hence we conclude that $W$ is non-decreasing
along the flow towards smaller values of $r$.  We also note that
the lhs of (\ref{c}) is nothing but $\ft13( {R_r}^r - {R_t}^t)$, which
equals $\ft13 ( {T_r}^r - {T_t}^t)$ through the Einstein equations 
\cite{FGPW}.  Equation (\ref{c}) then translates into 
 $ {T_r}^r - {T_t}^t \leq 0$, which yields 
the weaker positive energy condition \cite{FGPW}.  
Thus, in the presence
of a four-dimensional cosmological constant on the wall, 
$C(r)
\propto W^{-3}$ continues to play the role of a c-function,
i.e. of a function that is non-increasing along the flow towards the
infrared \cite{AG,FGPW}.

Using the above gravitational description of an RG flow, the
associated beta-functions are given by 
\bea
\beta^{\Lambda} = \mu
\frac{\partial \phi^{\Lambda}}{\partial \mu} =  \gamma^{-1} W^{-1}
\phi^{\prime \Lambda } \; ,
\label{beta}
\eea
where $\Lambda$ runs over both the vector and the hyper scalars.
The field theory has a conformal fixed point
whenever $\phi^{ \prime \Lambda } =0$.  Inspection of (\ref{floweqs}) 
then shows that $\partial_x W = \partial_X W=0$ at a fixed point.
Near a fixed point $\phi^{\Lambda}_{fix}$ we can write the BPS flow 
equations (\ref{floweqs}) 
for the scalar fields as $ (\delta \phi^{\Lambda})^\prime
= M^{\Lambda}\,_{\Sigma} \, \delta \phi^{\Sigma}$, where 
$\delta \phi^{\Sigma} = \phi^{\Sigma} - \phi^{\Sigma}_{fix}$.  Here
$M^{\Lambda}\,_{\Sigma}$ denotes a constant matrix with finite eigenvalues.
Hence we conclude that a fixed point $\phi^{\Lambda}_{fix}$ may only be
reached as $|r| \rightarrow \infty$.  For fixed scalars, the 
curved BPS flow equation
for $U$ can be solved, resulting in ${\rm e}^U = \Lambda (gW)^{-1}
\cosh (gW (r- r_0))$ \cite{DeWolfe:2000cp}.  As $|r| \rightarrow \infty$
we then obtain that ${\rm e}^U \rightarrow \infty$.  Hence we conclude
that for curved domain walls the fixed points 
can only be approached in UV-like directions
($\mu={\rm e}^U \rightarrow \infty, \gamma \rightarrow 1$).
This is in accordance
with the expectation that turning on a cosmological constant on the
wall does not affect the fixed-point behaviour of the dual field theory in
the UV.  On the other hand, as $\gamma \rightarrow 0$, 
we see from (\ref{beta}) and (\ref{floweqs}) that in general 
(some of) the beta-functions will blow up.

Finally, let us briefly comment on the energy functional (\ref{energy}).
When computing it on a curved BPS domain
wall solution, one finds that the boundary terms as well as the last
term in (\ref{energy}) are divergent at $r = \infty$.
In the case that 
there is a fixed point there, i.e. $W =
{\rm constant}$ and $U = g W r $ at $r = \infty$, these divergences
may be removed by boundary counterterms of the form $ W \sqrt{|\det g_{mn}|}$
and $W^{-1} \sqrt{|\det g_{mn}|} R$, where $R$ denotes the intrinsic 
curvature tensor computed from the induced metric $g_{mn}= 
{\rm e}^{2U} {\hat g}_{mn}$ \cite{HS,BK}.

\section{An example \label{secexa}}

Let us now explicitly construct a curved domain wall solution in a specific
gauged supergravity model
based on the coupling of the universal hypermultiplet to supergravity.
The associated quaternionic K\"ahler space is given by $\frac{SU(2,1)}{SU(2)
\times U(1)}$.  This space can be parametrized by the four hyper scalars
$q^X=(V,\sigma,\theta,\tau )$, and in this parametrization 
its metric is given by
$ds^2 = \ft12 V^{-2} dV^2 
+ \ft12 V^{-2} (d \sigma -2 \tau d\theta + 2 \theta d\tau)^2
+ 2 V^{-1} (d \theta^2 + d\tau^2) $.
We perform the gauging of the isometry
associated with the $U(1)$ transformation $C \rightarrow {\rm e}^{i \phi} C$,
where $C=\theta - i \tau$.  This $U(1)$ is part of the isotropy
group of the quaternionic 
manifold and therefore its gauging gives rise to a $W$ with a 
critical point \cite{CDKV}.
The associated triplet of Killing prepotentials
$P^s$ is given by \cite{CDKV} 
\bea
P^s=\sqrt{6} \left(- \frac{\theta}{\sqrt{V}},
- \frac{\tau}{\sqrt{V}}, \frac{1}{2} - \frac{\theta^2 + \tau^2}{2V} \right) \;.
\eea
Then $W = 1 + \frac{\theta^2 + \tau^2}{V}$ and 
\bea
Q^s = \left( -  \frac{2\sqrt{V} \theta}{V + (\theta^2 + \tau^2)},
-  \frac{2\sqrt{V} \tau}{V + (\theta^2 + \tau^2)}
, \frac{V - (\theta^2 + \tau^2)}{V + (\theta^2 + \tau^2)}\right) \;.
\eea
Observe that $W$ has a critical point at $\theta = \tau = 0$ ($\partial_X
W =0$).

In the flat case ($\gamma=1, B=0$), a solution to the BPS flow
equations (\ref{floweqs}) is given by
\bea
\sigma_0 = \theta_0 = 0 \;,\; \tau_0 = \sqrt{1-V_0} \;,\;
 V_0 =  1- {\rm e}^{-6r} \;,\;
U_0 = \frac{1}{6} \log({\rm e}^{6r} - 1) \;,
\label{flatsol}
\eea
where we set $g=1$ and where we picked the upper sign in (\ref{floweqs}).
The above trivially satisfies the equations of motion (\ref{moh})
for the hyper scalars.
We note that
this flat solution is supported by one hyper scalar field, 
since $V_0 + \tau_0^2 =1$.  This solution possesses a fixed-point at $r
\rightarrow \infty$ where $U_0 \rightarrow r$, and a curvature singularity
at $r=0$ where $U_0 \approx (6r)^{1/6}$.  
The range of $r$ is thus restricted to $0 < r < \infty$. 

Now consider perturbing this flat domain wall solution by turning
on curvature on the domain wall.  In order to solve the curved
BPS flow equations (\ref{floweqs}), we need to specify the triplet
$M^s$ which enters the definition of ${\Sigma_X}^Y$.  Let us introduce the
vector 
\bea
\vec{\omega} = \left(0, 
\frac{Q^3}{\sqrt{(Q^2)^2 + (Q^3)^2 }} , - 
\frac{Q^2}{\sqrt{(Q^2)^2 + (Q^3)^2 }} \right) \;,
\eea
which satisfies $\vec{\omega} \cdot
\vec{\omega} =1$ and $ \vec{\omega} \cdot \vec{Q} =0$.  We then take $\vec{M}$
to be given by $\vec{M} = \vec{\omega}$.  For this choice of $\vec{M}$, 
we can consistently truncate the model by setting $\sigma = \theta = 0$.  In 
doing so, not only do we find that the BPS equations
for $\sigma$ and $\theta$ are automatically
solved, but also the equations of motion
(\ref{moh}) for {\it all} the hyper scalars are identically
satisfied (to all orders in $|l|^{-1}$)!

Having thus checked the equations of motion, we return to the curved
BPS flow equations for the remaining fields $U, V$ and $\tau$, which are
given by
\bea
U^\prime &=& \gamma \, (1 + f)\, {V}^{-1} \;, \nonumber\\
V^\prime &=& 6 \gamma \, (1+f-V) + 12  |l|^{-1} \, 
{\rm e}^{-U} \sqrt{1+f-V} \,\frac{V^{3/2}}{1+f} 
\;, \nonumber\\
f^{\prime}  &=&  12  |l|^{-1}\, {\rm e}^{-U} 
\sqrt{1 + f -V} \,\sqrt{V} \;,
\eea
where we set $\tau = \sqrt{1+f-V }$.  Then, introducing the 
combinations 
\bea
T &=& {\rm e}^{-U} V (1 + f)^{-1} \;, \nonumber\\
h &=& V ( 1+f)^{-1} \;,
\eea
we obtain that $\gamma = \sqrt{1 - 4 |l|^{-2} T^2}$ as well as 
$T^{\prime} = \gamma \, T (5 h^{-1} -6)$ and $h^{\prime} = 6 \gamma \, (1-h)$.
{From} these equations we infer that
\bea
\frac{T^\prime}{T} = \frac{h^\prime \, (5-6 h)}{6 h (1-h)} \;.
\label{difth}
\eea
Integrating (\ref{difth}) yields 
\bea
T = h^{5/6} (1-h)^{1/6} \;, 
\label{th}
\eea
where we fixed the
integration constant by demanding that (\ref{th}) reduces to the appropriate
expression for the flat solution (\ref{flatsol}).  Thus we obtain that
\bea
{\rm e}^{-U} = \frac{T}{h} = \Big( h^{-1} - 1 \Big)^{1/6} \;,
\label{ul}
\eea 
where $h(r)$ satisfies the following differential equation 
by virtue of the various relations derived above:
\bea
h^\prime = 6 (1-h) \sqrt{1-4 |l|^{-2} \, h^{5/3} (1-h)^{1/3}} \;.
\label{hl}
\eea
The differential equation for $f$ may be
rewritten as follows:
\bea
\Big( \log (1+f) \Big)'  = 12 |l|^{-1} h^{1/3} (1-h)^{2/3} \;.
\label{fl}
\eea
Equations (\ref{ul})-(\ref{fl}) determine the curved BPS domain
wall solution to all orders in the cosmological constant.  
The function $h(r)$ can be obtained by numerical integration of (\ref{hl}).
Inserting it into (\ref{ul}) and (\ref{fl}) then yields $U(r), f(r)$
as well as $V(r) = h(r) (1 + f(r))$ and
$\tau (r) = \sqrt{(1-h(r)) (1 + f(r))}$.  
Thus, in contrast to the flat solution
(\ref{flatsol}), the curved solution is supported by two hyper scalar
fields.  

\begin{figure}
\begin{center}
\leavevmode \epsfxsize=7cm
 \epsfbox{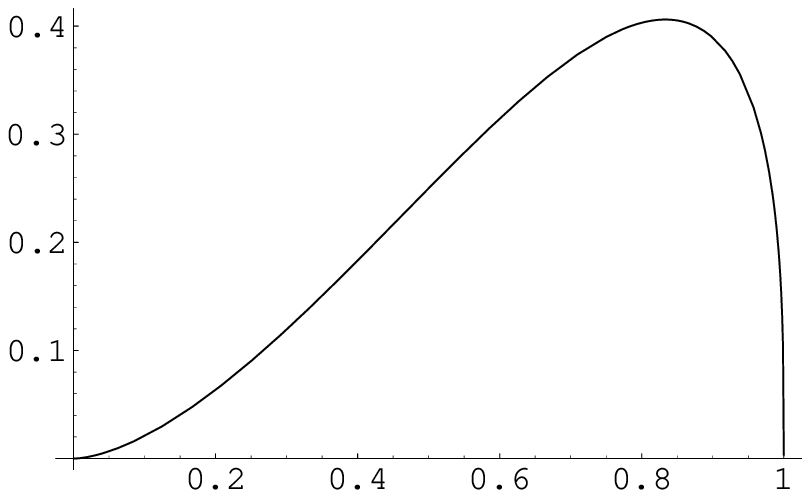} \hspace{1cm}\epsfxsize=7cm
 \epsfbox{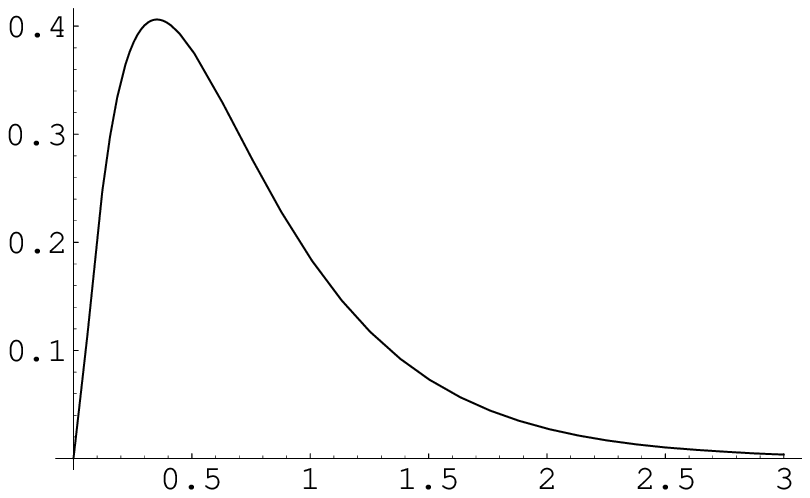}
\caption{ Left: $T^2 (h)$. Right: $T^2(r)$ for the value
$\Lambda =1$.  \label{fig:t2h}}
\end{center}
\end{figure}

Inspection of the relation $T^6 =h^5 ( 1-h)$  shows that $h(r)$ has the range
$0 \leq h \leq 1$.  {From} $h^{\prime} = 6 \gamma \, (1-h)$ we then infer that
$h(r)$ is a monotonic function in $r$, and hence also ${\rm e}^{U(r)}$.
The curved solution possesses the same fixed point and singularity structure
as the flat solution (\ref{flatsol}).  Namely, 
the point $h=1$ corresponds to $f={\rm const}, V={\rm const}, \tau =0, 
{\rm e}^U = \infty$, and hence to a fixed point at $r = \infty$.  The point
$h=0$, on the other hand, corresponds to a curvature singularity at $r=0$,
where $h \approx 6r$ and  $U \approx (6r)^{1/6}$. Observe that $\gamma =1$
at both $h=0$ and $h=1$.

\begin{figure}
\begin{center}
\leavevmode \epsfxsize=7cm
 \epsfbox{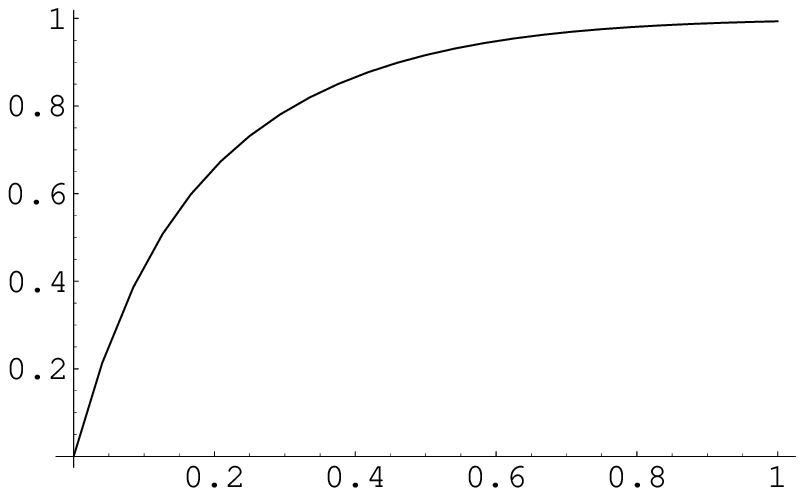}\hspace{1cm}\epsfxsize=7cm
 \epsfbox{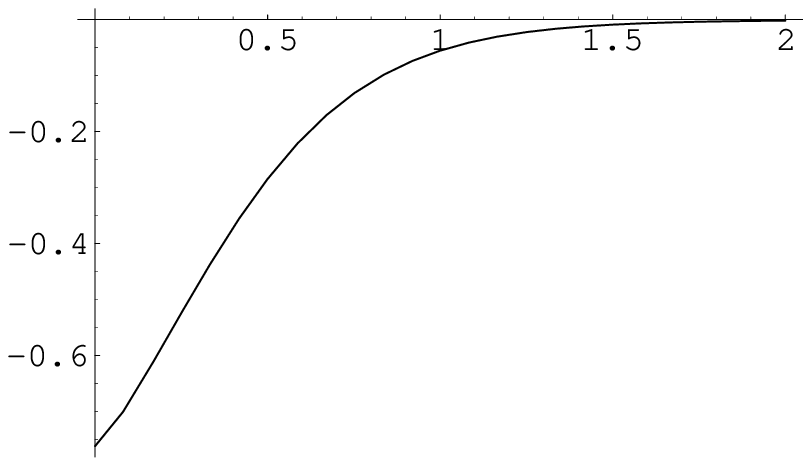}
\caption{
$h(r)$ (left) and $f(r)$
(right) for the value $\Lambda =1$. \label{fig:hf}}
\end{center}
\end{figure}

\begin{figure}
\begin{center}
\leavevmode \epsfxsize=7cm
 \epsfbox{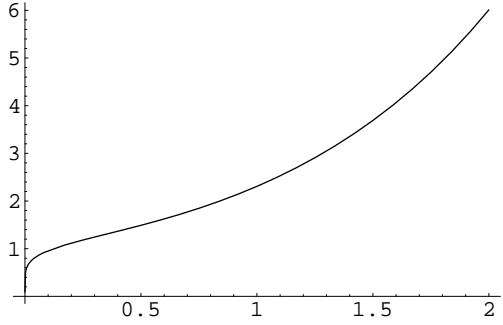}
\caption{ ${\rm e}^{U(r)}$ 
for the value $\Lambda =1$. \label{fig:u}}
\end{center}
\end{figure}

Using (\ref{th}), we can plot
$T^2$ over $h$. We see that $T^2$ has a maximum whose value
doesn't exceed $0.41$ (see Figure \ref{fig:t2h}).  
Hence it follows that for a small
value of $\Lambda^2 = 4 |l|^{-2}$ the function $\gamma = \sqrt{1-\Lambda^2
T^2}$ is positive, $\gamma > 0$, and the curved BPS solution exists
for the entire range $0<r<\infty$.  
It can be obtained by numerical integration
of (\ref{hl}) and (\ref{fl}).
In Figures \ref{fig:t2h}, 
\ref{fig:hf} and \ref{fig:u}
we have plotted $T^2(r), h(r), f(r)$ 
and ${\rm e}^{U(r)}$ for the value
$\Lambda =1$.

\begin{figure}
\begin{center}
\begin{picture}
(220,110)(0,0)
\put(0,0)
{\leavevmode \epsfxsize=7cm
 \epsfbox{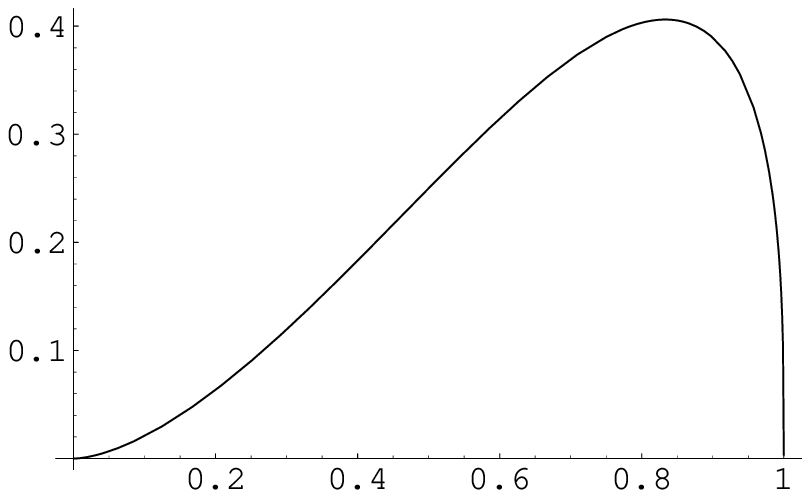} }
\put(100,80)
{\line(1,0){100}}
\put(145,83){$\Lambda^{-2} $}
\end{picture}
\caption{The two real solutions of $h^5 -h^6 = \Lambda^{-6}$ occur where
the horizontal line $\Lambda^{-2}$ intersects $T^2(h)$.
\label{fig:ths}}
\end{center}
\end{figure}

When increasing the value of 
$\Lambda^2$, there is a critical value $\Lambda^2_c$ at which 
$\Lambda^{-2}_c$ equals the maximum of $T^2(h)$ ($T^2_{\rm max} =
\Lambda^{-2}_c $).  Thus there is a particular point $r=r_c$ where
$\gamma (r_c) =0$.  Since $\gamma (r) \geq 0$ over the range
$0<r<\infty$, there is yet no obstruction to having a curved
solution over the entire range of $r$.  However, when continuing
to increase the value of $\Lambda^2$ ($\Lambda^2 > \Lambda^2_c$), the
maximum of $T^2(h)$ becomes much larger than $\Lambda^{-2}$.  Then $\gamma$
develops two zeros determined by the two real solutions of the
equation $h^5 - h^6 = \Lambda^{-6}$ (see Figure \ref{fig:ths}).  
Let us denote these two solutions
by $h_I$ and $h_{II}$, where $h_I < h_{II}$.  They occur at two specific
values for $r$, $r_I$ and $r_{II}$, with $r_I < r_{II}$.  Thus, in the
region $h_I < h < h_{II}$ (corresponding to $r_I < r < r_{II}$) $\gamma$
is imaginary, and there does not exist a real curved solution to the BPS flow
equations.  The curved domain wall
solution only exists in the regions $0 < h \leq h_I$
and $h_{II} \leq h < 1$ corresponding to $0<r \leq r_I$ and $r_{II} \leq r<
\infty$, respectively.

The curved BPS solution constructed above made use of a specific choice
for $\vec{M}$, namely $\vec{M} = \vec{\omega}$.  In principle
there may exist other choices for $\vec{M}$ which also lead to curved
BPS solutions.  Let us now pick a more general $\vec{M}$, namely
$ \vec{M} = c(r) \vec{\omega} + 
\sqrt{1 - c^2(r)} \, \vec{\omega} \times \vec{Q}$. 
When evaluated on the flat solution (\ref{flatsol}), $\vec{M}$ is given by 
\bea
\vec{M} 
= \Big(\sqrt{1-c^2(r)},c(r)\,(-1+2V_0), 2 c(r)\, \sqrt{1-V_0} \sqrt{V_0}
\Big) \;.
\eea
For this more general $\vec{M}$,
and in contrast with the previous choice, we are only
able to construct the curved domain wall solution order by order in 
$|l|^{-1}$.  To be specific, 
let us then determine the 
lowest order modification of the flat solution.
Since $B = 2 |l|^{-1} {\rm e}^{-U} 
W^{-1}$, the lowest order modification is of order $|l|^{-1}$.  At this order,
$B=  2 |l|^{-1} {\rm e}^{-U_0} V_0$ and $\gamma =1$.  
We now determine the proportionality function $c(r)$ by demanding 
that the equations of motion (\ref{moh})
for the hyper scalars be satisfied to order $|l|^{-1}$.
Observe that we may then use the flat domain wall solution in (\ref{moh}),
since to order $|l|^{-1}$ each term comes explicitly multiplied by  $|l|^{-1}$.
We find that to
order
$|l|^{-1}$ all terms in (\ref{moh}) cancel out provided that $c$ is constant!

Having determined $M^s$ to order $|l|^{-1}$, we now turn
to the BPS flow
equations (\ref{floweqs}).  
We again set
$\tau = \sqrt{1+f-V }$.
Expanding $V$ and $U$ as $V=V_0 + 
V_1$ and $U = U_0 + U_1$,
we find that to order $|l|^{-1}$ the BPS flow equations are solved by
\bea
U_1 &=& 0 \;,
\nonumber\\
V_1 &=& - 
c\,|l|^{-1} (1-{\rm e}^{-6r}) \, H(r)
\;, \nonumber\\
\sigma &=& 
\sqrt{1-c^2} \,|l|^{-1}  H(r) 
\;,
\nonumber\\
\theta &=& 
\frac{1}{2} 
\sqrt{1-c^2} \,|l|^{-1}   {\rm e}^{-3r} \, H(r) 
\;, \nonumber\\
f&=& - 
c \,|l|^{-1} H(r) 
\;, 
\label{solfirst}
\eea
where
\bea
H(r) = 
2 {\rm e}^{-4r} (1-{\rm e}^{-6r})^{1/3} + {\rm e}^{-4r}\,
{\rm {_2}F_1}\left(\frac{2}{3}, \frac{2}{3}, \frac{5}{3}, {\rm e}^{-6r}
\right) \;, 
\eea
and where we set various integration constants to zero.
Thus, to lowest order in $|l|^{-1}$, (\ref{solfirst}) (together with
(\ref{flatsol})) describes
a curved domain
wall solution that solves both the BPS flow equations and the 
equations of motion.  This  curved 
solution is supported by
at least two hyper scalar fields. In the case that $c=1$, the curved
solution  (\ref{solfirst}) is in agreement with (\ref{ul})-(\ref{fl})
when expanded to lowest order in $|l|^{-1}$.

As a further application of our results, 
it would be interesting to construct a curved version
of the flow discussed in \cite{CDKV}.  This flow, which is an 
${\cal N}=2$ version of the FGPW flow \cite{FGPW}, is
described by a flat wall 
which is supported
by one vector scalar and one hyper scalar.  
We observe that the triplet $M^s$, which is necessary for the construction 
of the curved wall, can now 
be determined via (\ref{pq}). Yet another interesting application
would be to construct a curved version of the GPPZ flow
\cite{GPPZ2}.  We hope to return to these issues in the future.

\bigskip

{\bf Acknowledgements}

We would like to thank  L. Alvarez-Gaum\'e, 
K. Behrndt, G. Curio and M. 
Strassler
for valuable discussions.
This work
is supported by the DFG and by the European Commission RTN Programme
HPRN-CT-2000-00131. 





\providecommand{\href}[2]{#2}\begingroup\raggedright

\endgroup

\end{document}